\documentclass[useAMS,usenatbib]{mn2e}
\usepackage{graphicx}
\usepackage{rotating,times,pictex,graphicx,latexsym}
\usepackage{color}
\usepackage{longtable,amsmath}
\usepackage{lscape}

\title{FSR\,0190 -- Another old distant galactic cluster} 

\author[Froebrich, Meusinger \& Davis]{D.~Froebrich$^{1}$\thanks{E-mail:
df@star.kent.ac.uk}, H.~Meusinger$^{2}$ and C.J.~Davis$^{3}$\\ $^1$ Centre for
Astrophysics and Planetary Science, University of Kent, Canterbury, CT2 7NH, UK
\\ $^2$ Th\"uringer Landessternwarte Tautenburg, Sternwarte 5, 07778 Tautenburg,
Germany \\ $^3$ Joint Astronomy Centre, 660 North A\`{}ohoku Place, University
Park, Hilo, Hawaii 96720, USA } 

\begin{document}

\date{Received sooner; accepted later}
\pagerange{\pageref{firstpage}--\pageref{lastpage}} \pubyear{2007}
\maketitle

\label{firstpage}

\begin{abstract}

We are conducting a large program to classify newly discovered Milky Way star
cluster candidates from Froebrich et al. \cite{2007MNRAS.374..399F}. Here we
present NIR follow-up observations of FSR\,0190 ($\alpha$=$20^{h} 05^{m}
31^{s}.3$, $\delta$=$33^{\circ} 34^{'} 09^{''}$ J2000). The cluster is situated
close to the Galactic Plane ($l$=$70.7302^{\circ}$, $b$=$+0.9498^{\circ}$). It
shows a circular shape, a relatively large number of core helium burning stars
-- which clearly distinguishes the cluster from the rich field -- but no
centrally condensed star density profile. We derive an age of more than 7\,Gyr,
a Galactocentric distance of 10.5\,kpc, a distance of 10\,kpc from the Sun, and
an extinction of $A_{\rm K}$\,=\,0.8\,mag. The estimated mass is at least of the
order of 10$^5$\,M$_\odot$, and the absolute brightness is $M_{\rm
V}$\,$\le$\,$-$4.7\,mag; both are rather typical properties for Palomar-type
globular clusters.

\end{abstract}


\section{Introduction}
 
A large number of stars form in clusters. Over time most clusters will dissolve
and the cluster stars will migrate into the field, e.g. due to tidal
interactions with giant molecular clouds. The study of the distribution of old
clusters in the Galaxy will hence shed light on the disruption time-scales and
the underlying physical processes. Furthermore, galactic globular clusters
(GlCl) allow us to probe the conditions during the time of the formation of our
Galaxy. Pre-requisites for these investigations are large, well defined samples
of objects. Currently the sample of known old open clusters is very incomplete
(e.g. Bonatto \& Bica \cite{2007MNRAS.in.pressB}). Also the number of newly
discovered galactic GlCls in recent years (see Bonatto et al.
\cite{2007MNRAS.in.pressB2} or Bica et al. \cite{2007A&A...472..483B} for a
summary) suggests that this sample as well is incomplete, especially at the low
mass/luminosity end -- the Palomar-type GlCls. 

Wide field searches at infrared wavelength have provided a wealth of new
clusters and candidates in recent years. Based on star counts in 2MASS
(Skrutskie et al. \cite{2006AJ....131.1163S}),  Froebrich et al.
\cite{2007MNRAS.374..399F} presented a list of 1021 new cluster candidates. To
improve the usefulness of this sample (contamination rate about 50\,\%) a
classification of the clusters based on e.g. infrared colour-magnitude diagrams
has to be performed. One expects that the majority of these cluster candidates
are young embedded clusters. However, so far also three old (age
$\approx$\,1Gyr) open clusters (FSR\,0031/0089/1744, Bonatto \& Bica
\cite{2007MNRAS.in.pressB}) and three GlCls (FSR\,0584, Bica et al.
\cite{2007A&A...472..483B}; FSR\,1735, Froebrich et al.
\cite{2007MNRAS.377L..54F}; FSR\,1767, Bonatto et al.
\cite{2007MNRAS.in.pressB2}) have been identified. In this work we present our
results on the classification of FSR\,0190 based on new deep near infrared
observations.

The paper is structured as follows. Our data is presented in Sect.\,\ref{data}
and results including the appearance of the cluster, the contamination with
field stars and the isochrone fitting to determine the cluster properties are
presented in Sect.\,\ref{results}. Finally in Sect.\,\ref{discussion} we discuss
and conclude our findings.

\section{Data}

\label{data}
 
We secured near-infrared (NIR) J, H, and K-band imaging data with UFTI (Roche et
al. \cite{2003SPIE.4841..901R}) at the U.K. Infrared Telescope (UKIRT) on the
29th of May, 2007. We obtained a 5$\times$5 frame mosaic mapped with half
detector spacing to cover the entire field of the cluster. The pixel size in the
images is 0.09\arcsec. The data where taken under photometric conditions with
120\,sec per pixel integration time in each filter. Standard data reduction
techniques (dark subtraction  and self-flat-fielding) before image-registration
and mosaicking  was performed using the facility pipeline {\em ORAC-DR}
(Cavanagh et al. \cite{2003ASPC..295..237C}).

Despite the excellent seeing conditions (stellar FWHM are about 0.55\arcsec\, in
the final JHK mosaics), our photometry suffers from crowding in the field, due
to the cluster's position close to the Galactic Plane
($l$\,$\approx$\,70.73$^{\circ}$, $b$\,$\approx$\,$+$0.95$^{\circ}$). Using 
3\,$\sigma$ detections in the K-band image, there are about 6500 stars in the
field. If each star occupies an area with a diameter equal to the seeing (the
minimum area required for photometry), then there is a crowding of 3\,\% in the
field. This increases to 4\,\% when using 2\,$\sigma$ detections. We have
performed our photometry using the SExtractor software (Bertin \& Arnouts
\cite{1996A&AS..117..393B}). Only 2\,$\sigma$ detections in the K-band, with
quality flags better than 3 and photometric errors below 0.2\,mag in all three
filters are used in our subsequent analysis. 

We have used the 2MASS point sources in the field to flux calibrate our JHK
images. The $rms$ scatter in the calibration is 0.14, 0.13, 0.10\,mag in JHK
respectively caused mainly by magnitude migration towards brighter magnitudes
due to  the strong image crowding in this field. Since the 2MASS data are based
on much lower resolution,  the scatter is expected to be dominated by the
uncertainties in the 2MASS magnitudes  of the calibration stars.

\section{Results}

\label{results}

\begin{figure}
\centering
\fbox{\includegraphics[width=8cm]{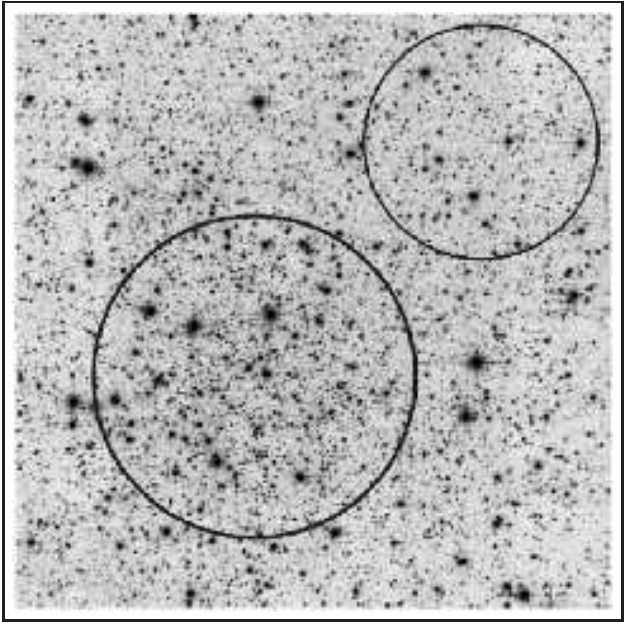}}

\caption{\label{kgray} Logarithmic K-band gray scale image of FSR\,0190. The
cluster is positioned off-center towards the south-east. The two circles
indicate the cluster area and the control field. The image is 4.35\arcmin
$\times$4.35\arcmin\, in size.} 

\end{figure}

\begin{figure*}
\centering
\hfill
\fbox{\includegraphics[width=7cm]{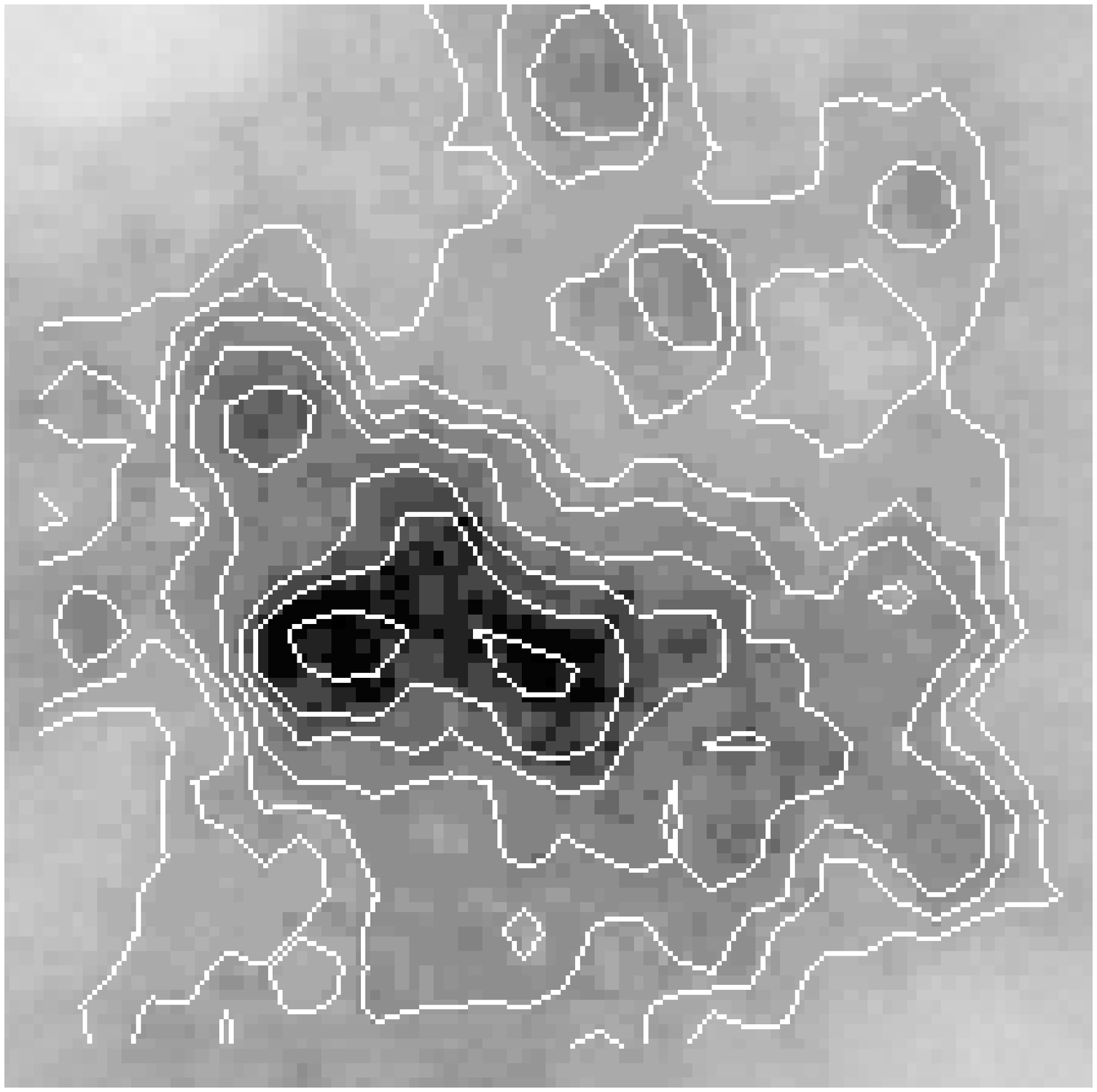}}\hfill
\fbox{\includegraphics[width=7cm]{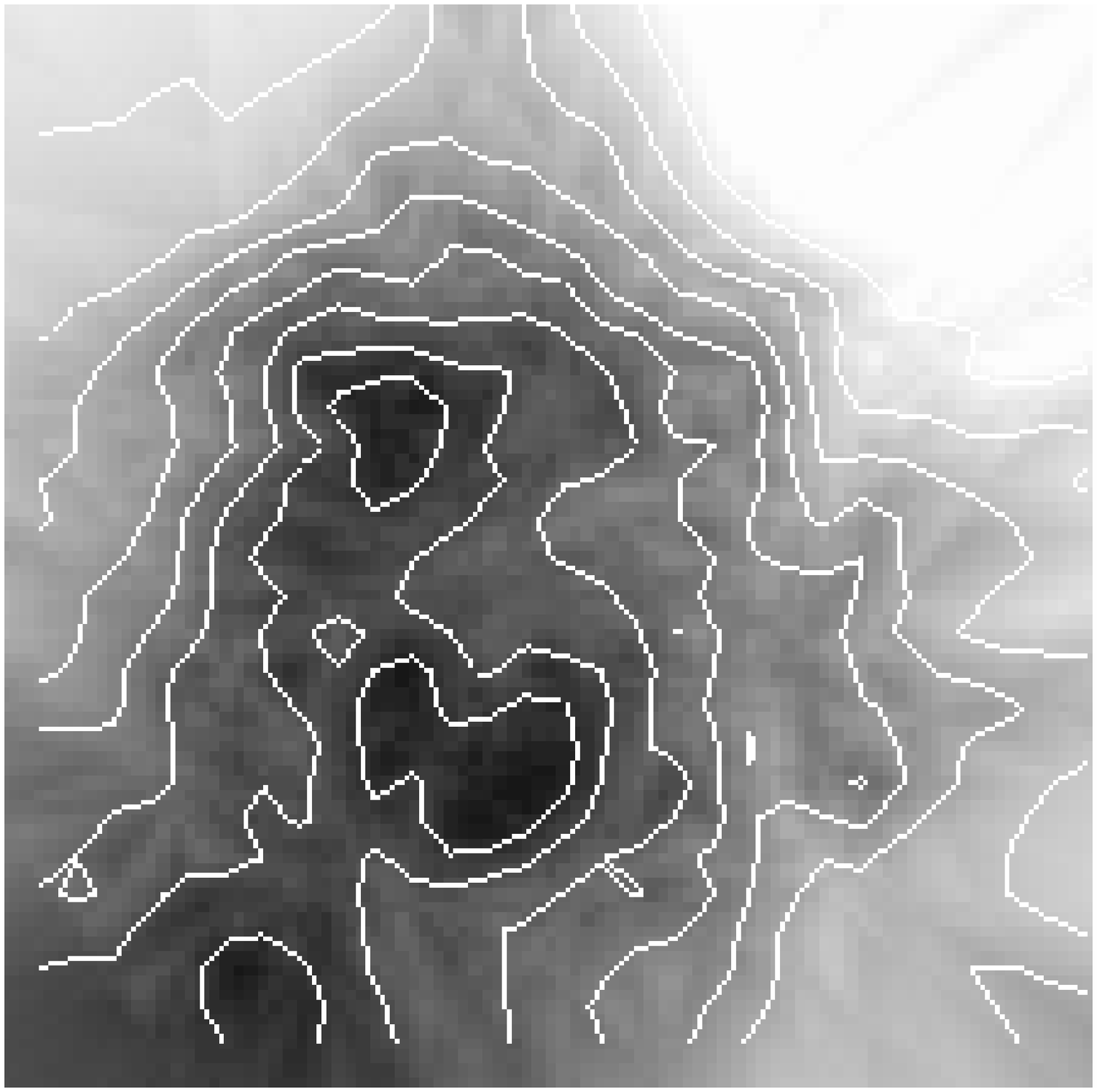}}\hfill
\hfill

\caption{\label{contour} {\bf Left:} Nearest neighbour plot for all detected
stars in the field. The gray-scale values indicate the distance to the 85th
closest neighbour star. Contours range from 23\arcsec\, to 17\arcsec\, in steps
of 1\arcsec. {\bf Right:} As in the left panel but only using core helium
burning stars (selection see text) and the distance to the 10th closest
neighbour. Contours range from 70\arcsec\, to 30\arcsec\, in steps of 5\arcsec.
The images are 4.35\arcmin $\times$4.35\arcmin\ in size.
} 

\end{figure*}

\subsection{Cluster Appearance}

\label{appearance}

We show the K-band mosaic of the field around the cluster in Fig.\,\ref{kgray}.
The cluster itself is positioned south-east of the centre. An increase in the
stellar density can be seen. Due to the close proximity of the Galactic Plane
the star density contrast between cluster and field is rather low. The cluster
can more clearly be identified in the map showing the distance to the 85th
nearest neighbour of each star (left panel in Fig.\,\ref{contour}). In this
panel one can easily identify a star density enhancement south-east of the image
centre. The average star density in the cluster area is a bit less than twice as
high as in the field/control area (upper right corner). Hence the star cluster
itself possesses a smaller star density as the field integrated along the line
of sight.  The cluster shows no centrally condensed appearance, but rather a
more or less uniformly increased star density across an extended region. A fit
of the radial star density by a King-profile results in $r_{\rm
core}$\,=\,65$\pm$8\arcsec\, and $r_{\rm tid}$\,=\,195$\pm$40\arcsec.

We have created a further nearest neighbour map using only stars that have
colours and magnitudes consistent with them being core helium burning objects
associated with the cluster (selection see below). In this case the contrast
between the cluster and the field becomes much more clear. There are about five
times more such stars per unit area in the cluster region than in the control
field. In the right panel of Fig.\,\ref{contour} we show the distance to the
10th nearest neighbour for all these stars. There is some sub-structure visible
in the map, indicating two regions (south and north of the cluster centre) where
the density of these objects is enhanced. It is not clear if this effect is
real, or just resembles a selection effect caused by our criteria for stars with
acceptable photometry. The same applies to the structure seen in the nearest
neighbour plot for all stars in the area.

\subsection{Field star decontamination}
\label{removal}

\begin{figure*}
\centering
\hfill
\includegraphics[width=6.1cm, angle=-90]{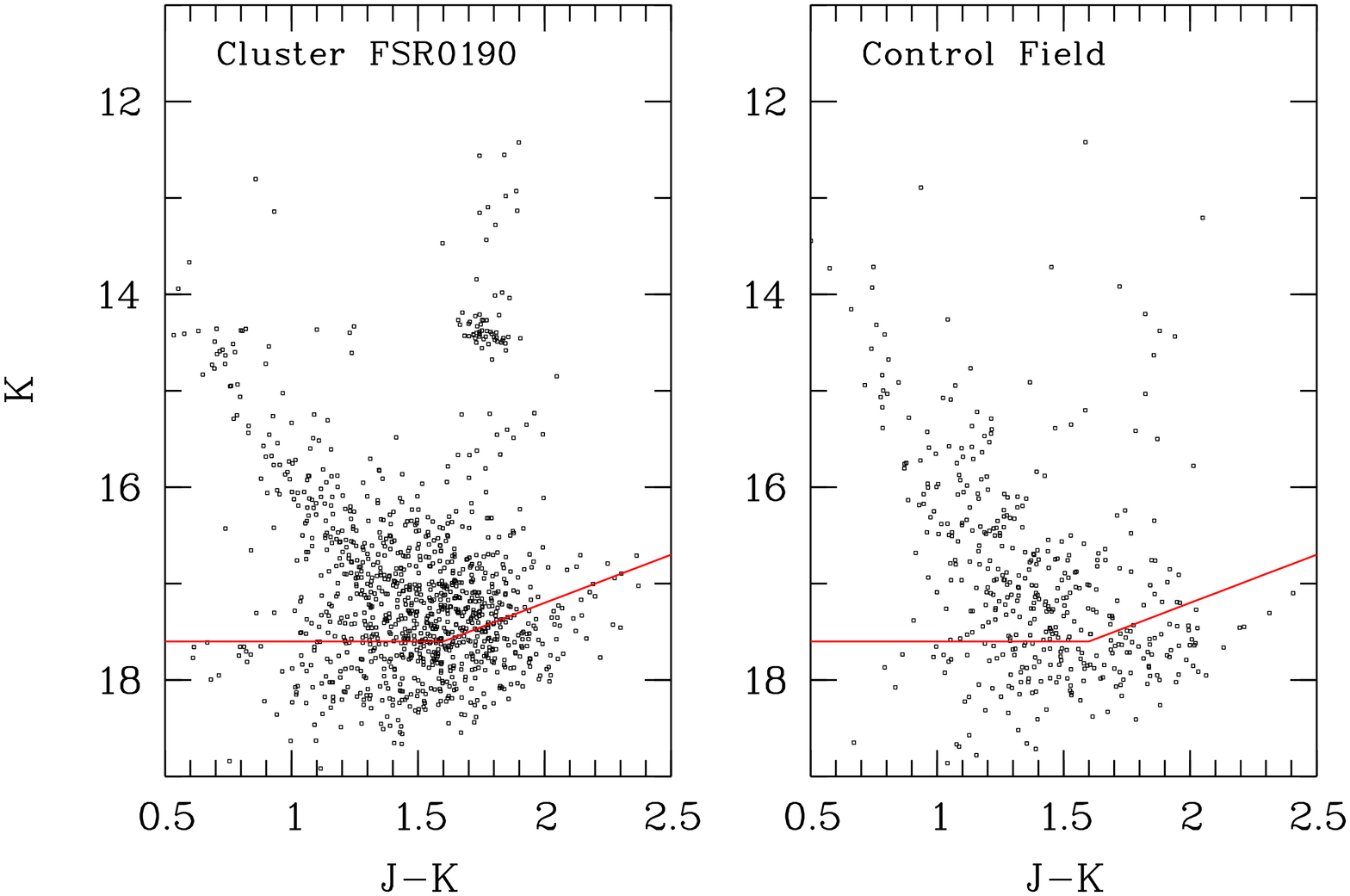} \hfill
\includegraphics[width=6.1cm, angle=-90]{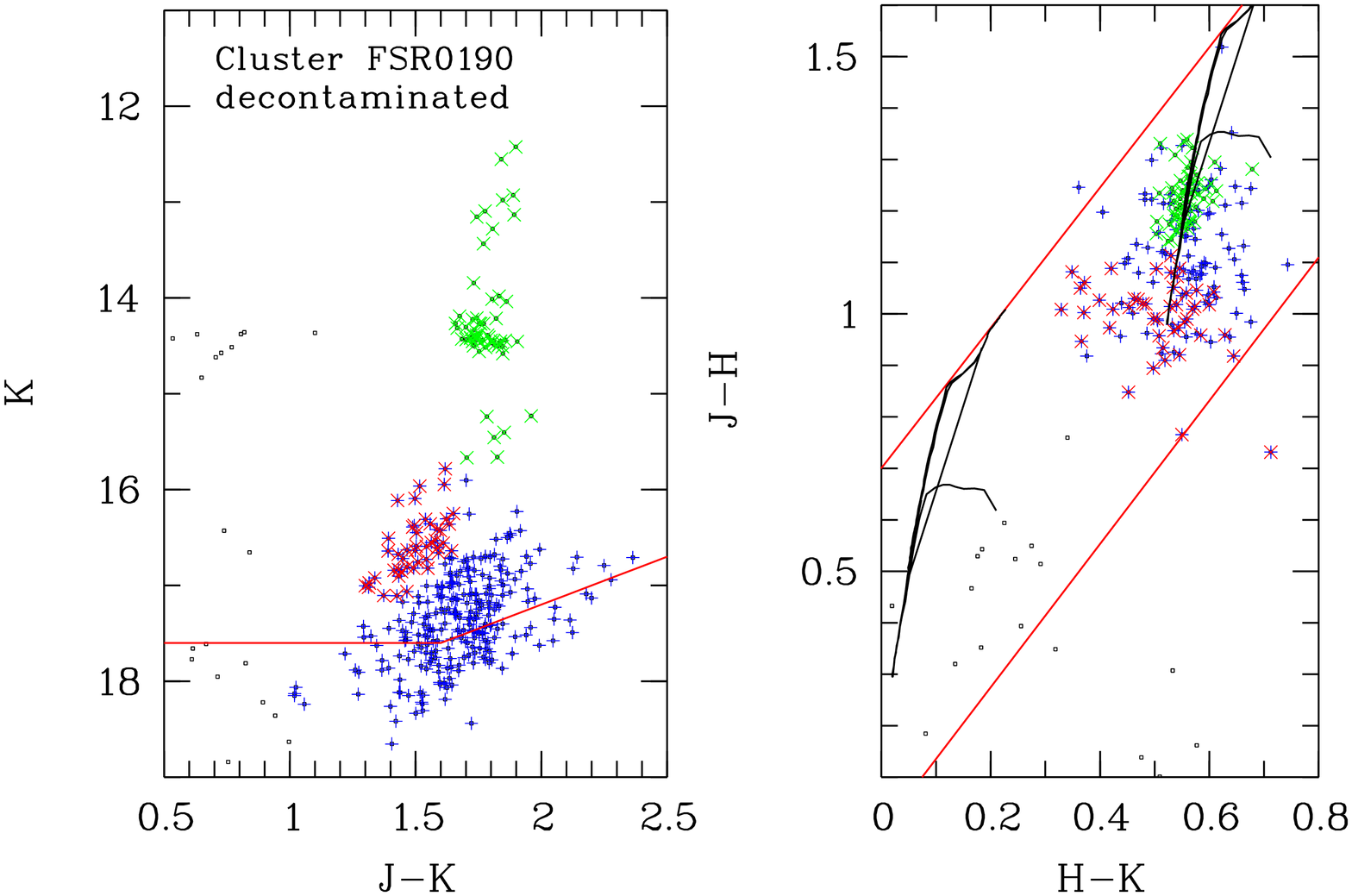}\hfill

\caption{\label{colmag} {\bf 1st panel:} J-K vs. K colour-magnitude diagram of
the cluster area (radius 72\arcsec). The solid line indicates the completeness
limit of our data. {\bf 2nd panel:} As in the 1st panel but for the control area
with a radius of 52\arcsec. {\bf 3rd panel:} One realisation of the
decontaminated J-K vs. K colour-magnitude diagram of the cluster area (radius
72\arcsec). Crosses indicate RGB/AGB, $+$ signs main sequence and asterisks
suspected blue straggler stars in the cluster. {\bf 4th panel:} The same stars
as in the 3rd panel in a H-K vs. J-H colour-colour diagram. The straight solid
lines indicate the reddening path for main sequence and giant stars, whose loci
from a fit of a 10\,Gyr and Z\,=\,0.004 isochrone (with and without reddening)
are over plotted. Only stars with photometric errors of less than 0.1\,mag in all
three filters are shown for clarity.} 

\end{figure*}

As discussed in Section\,\ref{appearance}, the population of field stars
contributes significantly to the stellar density. Hence the analysis of the
cluster properties is strongly hampered. This can be seen in the J-K vs. K
colour-magnitude diagrams in Fig.\,\ref{colmag}. There we compare all stars
detected in the area of the cluster (1st panel) and the control field (2nd
panel). The different number of stars can be explained by the different areas
covered by the cluster (4.52 square arcminutes) and the control field (2.38
square arcminutes). However, there are clearly differences in the two diagrams,
most notably the group of stars in the cluster area at about J-K\,=\,1.75\,mag
and K\,=\,14.4\,mag. 

We have used the colour and magnitude information of the stars in the control
field to statistically remove foreground and background stars from the cluster
area. In particular we adopted the decontamination algorithm described in
Bonatto \& Bica \cite{2007MNRAS.377.1301B}. We used cell dimensions of
$\Delta$J\,=\,0.5\,mag, $\Delta$(J-H)\,=\,0.2\,mag and
$\Delta$(J-K)\,=\,0.2\,mag to compute the expected number of field stars. This
number of field stars is then randomly removed from the objects in the cluster
area. In panel 3 of Fig.\,\ref{colmag} we show one such realisation of the
decontamination of the cluster area, while the 4th panel shows the
decontaminated H-K vs. J-H colour-colour diagram. Over plotted are the loci of
main sequence and giant stars from a fit of a 10\,Gyr and Z\,=\,0.004 isochrone
(see below), with and without reddening.

\begin{figure*}
\centering
\includegraphics[height=4.34cm, angle=-90, bb = 28 35 570 410]{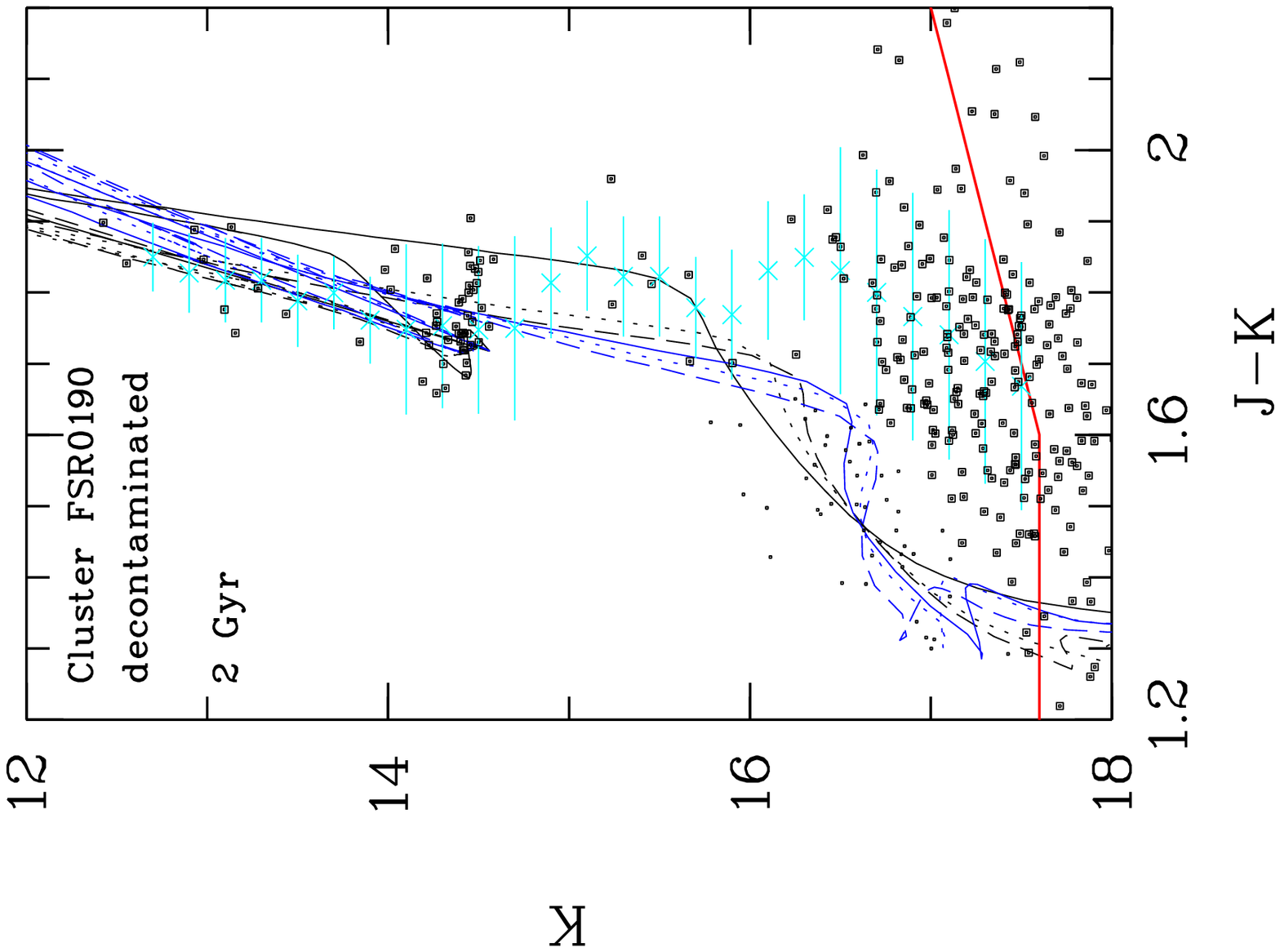} \hfill
\includegraphics[height=4.34cm, angle=-90, bb = 28 35 570 410]{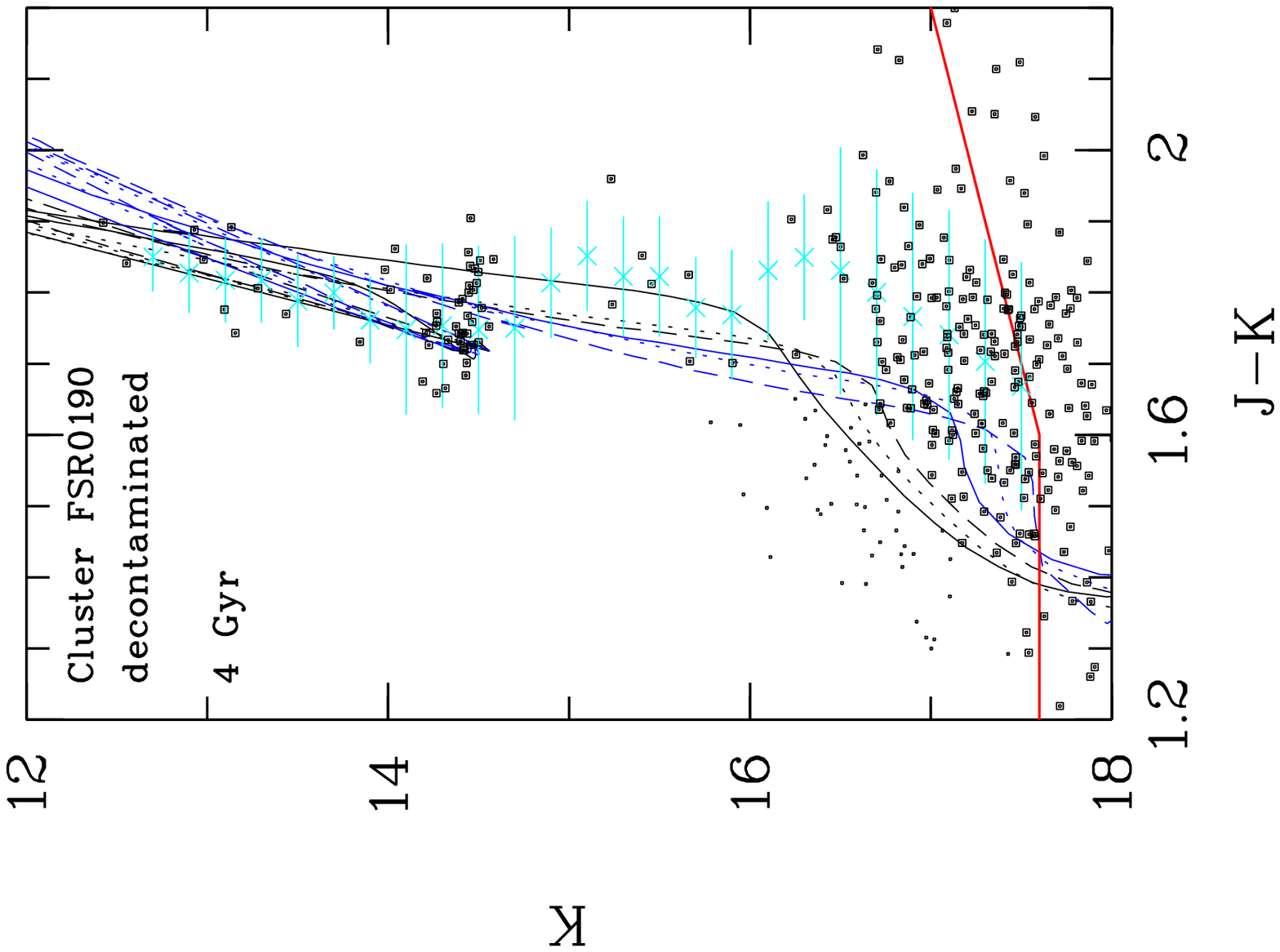} \hfill
\includegraphics[height=4.34cm, angle=-90, bb = 28 35 570 410]{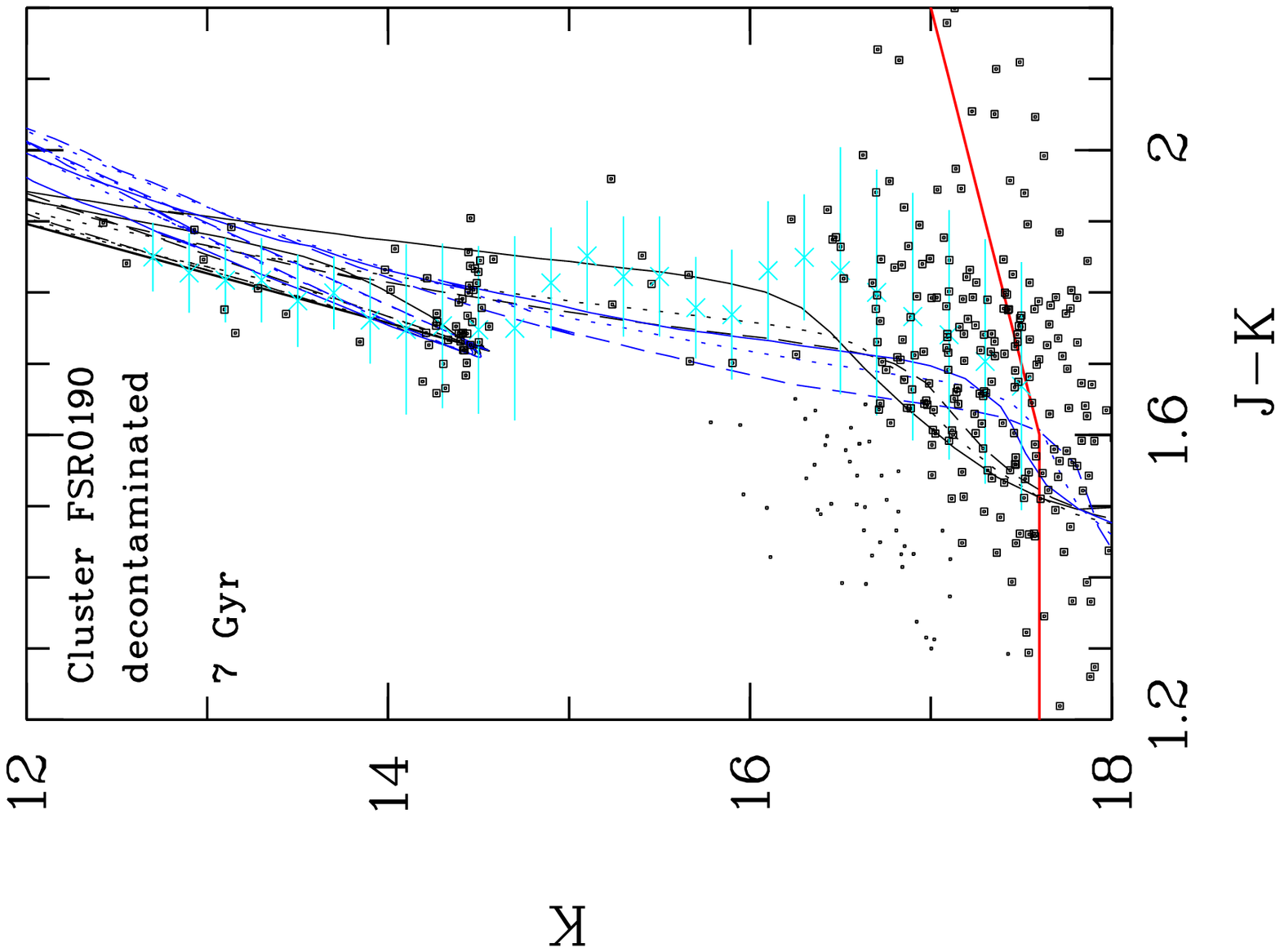} \hfill
\includegraphics[height=4.34cm, angle=-90, bb = 28 35 570 410]{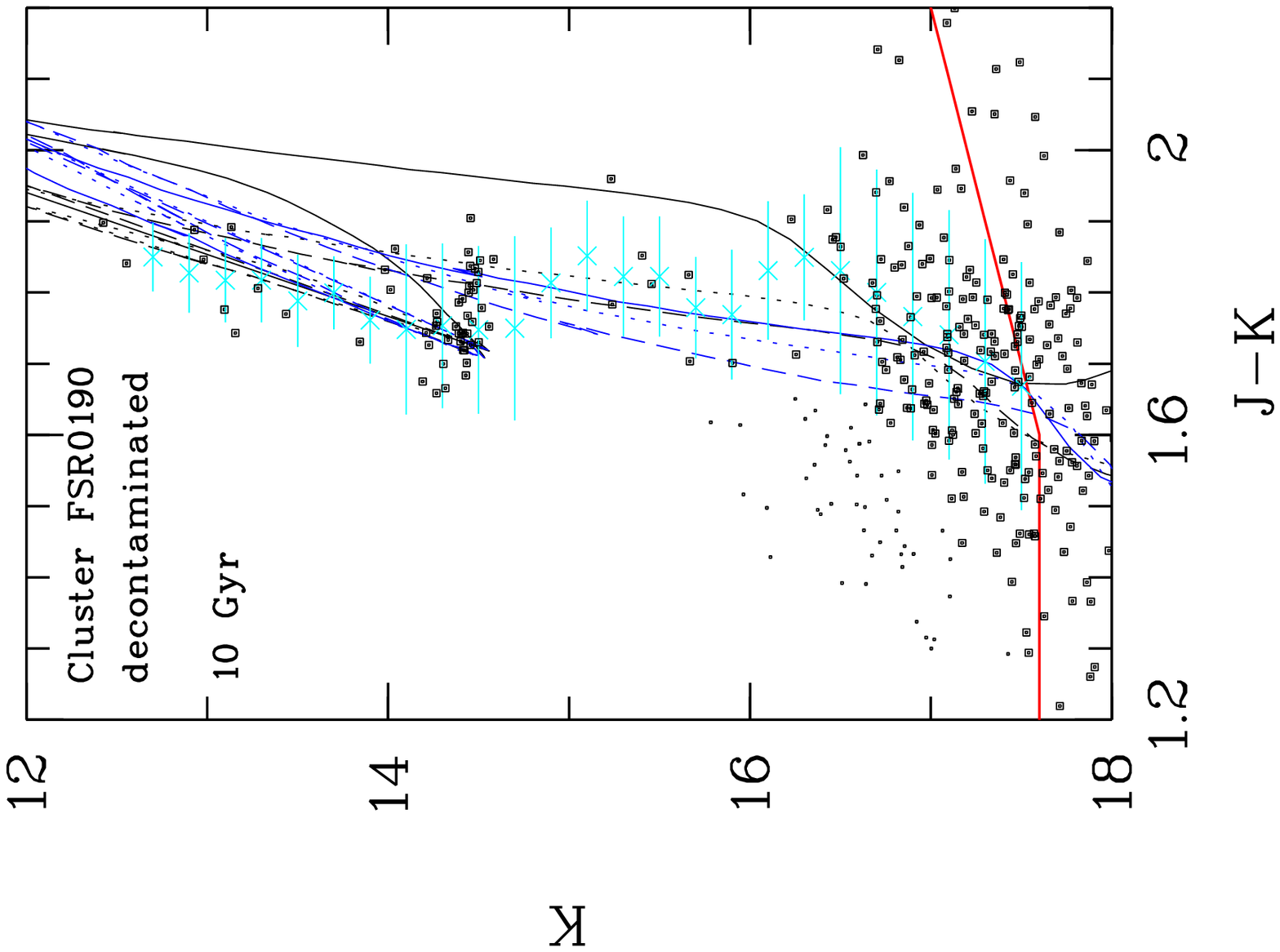} \\

\caption{\label{iso} Decontaminated colour-magnitude diagrams for the cluster
area. Over plotted are isochrones of different ages (indicated in each panel) and
metallicities ($Z$\,=\,0.0001 solid-dark; $Z$\,=\,0.0004 dotted-dark;
$Z$\,=\,0.001 dashed-dark; $Z$\,=\,0.004 solid-gray; $Z$\,=\,0.008 dotted-gray;
$Z$\,=\,0.019 dashed-gray). The light gray crosses and error bars indicate the
observed distribution of cluster stars (not considering the suspected blue
stragglers - small dots) and its scatter. The solid line at the bottom indicates
the completeness limit of the observations.} 

\end{figure*}

In the decontaminated cluster area we find the group of stars remaining at
around J-K\,=\,1.75\,mag and K\,=\,14.4\,mag. These are interpreted as the core
helium burning stars in the cluster. There are a number of brighter objects with
similar colours, most probably RGB/AGB stars. All these objects form a well
defined group in the colour-colour diagram. Furthermore, there are stars
remaining in the J-K vs. K diagram with colours of J-K\,$>$\,1.5\,mag and
K\,$>$\,16\,mag. This feature is identified with cluster main sequence stars
close to the main sequence turnoff. These stars occupy a region in the H-K vs.
J-H diagram that is consistent with this proposal. The larger scatter can be
explained by their lower magnitudes. In the J-K vs. K diagram these stars are
apparently split in two groups, one of which (about 40 stars) seems to be
off-set towards bluer colours and brighter magnitudes (3rd panel of
Fig.\,\ref{colmag}). We interpret this group as blue straggler stars (BSS) in
the cluster and the second, much larger, group as the cluster main sequence
turn-off. Note that some objects in the region of the suspected BSS might be
unresolved binaries, which have not been removed by the decontamination
procedure. The BSS interpretation is further supported by the following points:
i) If the small group is used as main sequence turn-off we cannot find an
isochrone (see Sect.\,\ref{fitting} and Fig.\,\ref{iso}) that fits all the
cluster stars, i.e. the majority of objects around K\,=\,17\,mag lack an
explanation. ii) The stars in the small group correspond to earlier spectral
types when plotted in the H-K vs. J-H diagram. iii) The number of BSS and their
homogeneous spatial distribution in the cluster are in agreement with the results
of Davies et al. \cite{2004MNRAS.349..129D}.

 %
 %

\subsection{Isochrone Fitting}

\label{fitting}

In order to constrain the basic cluster properties (age, metallicity, distance,
reddening) we fit isochrones to the decontaminated J-K vs. K colour magnitude
diagram. We used isochrones based on Girardi et al. \cite{2002A&A...391..195G}
computed for the 2MASS
filters\footnote{http://stev.oapd.inaf.it/~lgirardi/cgi-bin/cmd}. For each case
we matched the core helium burning stars in the isochrone and the observations
by varying the reddening E(J-K) and the distance of the cluster. In
Fig.\,\ref{iso} we show sets of isochrones with different ages for each panel
(2, 4, 7, 10\,Gyr). In each panel six isochrones are over plotted, using
different metallicities ($Z$\,=\,0.0001, 0.0004, 0.001, 0.004, 0.008, 0.019).
The panels also contain the observed distribution of cluster stars indicated by
light-gray crosses and error bars, as well as our completeness limit (determined
as the maximum of the K-band luminosity function). Note that we do not consider
the suspected BSS stars when comparing the isochrones with the cluster stars
(see below for some more detailed remarks).

From Fig.\,\ref{iso} it becomes apparent that isochrones with ages of 2 to
4\,Gyr do not fit the observed upper end of the main sequence. The same applies
also for all other lower ages. For ages above 7\,Gyr, most isochrones fit the
main sequence stars well, with maybe the exception of the lowest metallicities
($Z$\,=\,0.0001). In turn, the highest metallicity isochrones have difficulties
explaining the distribution of RGB/AGB stars brighter then the core helium
burning objects. The slope of the $Z$\,$>$\,0.008 isochrones in the
colour-magnitude diagram is too shallow compared to the observations. We
additionally can use the H-band information, i.e. match the isochrones with the
position of the core helium burning objects in the H-K vs. J-H diagram (see
right panel in Fig.\,\ref{colmag}). This requires a reddening of about $A_{\rm
K}$=0.8\,mag, and thus puts a tighter constraint on the metallicity, since the
fit of the isochrones in the J-K vs. K diagram has the extinction as a free
parameter. The metallicity range obtained from this process is
[M/H]\,=\,$-$0.9$\pm$0.4\,dex.

We have not used the suspected BSS stars in the comparison of the isochrones and
the observations. If this group is considered as the main sequence turn-off,
then the majority of stars below K\,=\,16\,mag in the cluster can not be
explained (see e.g. left panel of Fig.\,\ref{iso}). In case all the stars below
K\,=\,16\,mag are main sequence turn-off stars the cluster will have an age
lower than 7\,Gyr, but still in excess of 2\,Gyr.

Hence, we can constrain that the age of the cluster is larger than 7\,Gyr, while
the metallicity is most probably in the range of $Z$\,=\,0.001 to 0.006
([M/H]\,=\,$-$0.9$\pm$0.4\,dex). The average distance from the Sun required for
the fit is $r_\odot = 10.0$\,kpc, with an uncertainty of 1.0\,kpc, mostly caused
by the poorly constrained metallicity. The average values correspond to an age
of 10\,Gyr and $Z$\,=\,0.004. The cluster would have a smaller distance for
lower metallicities.

\begin{figure}
\centering
\includegraphics[width=6cm, angle=-90]{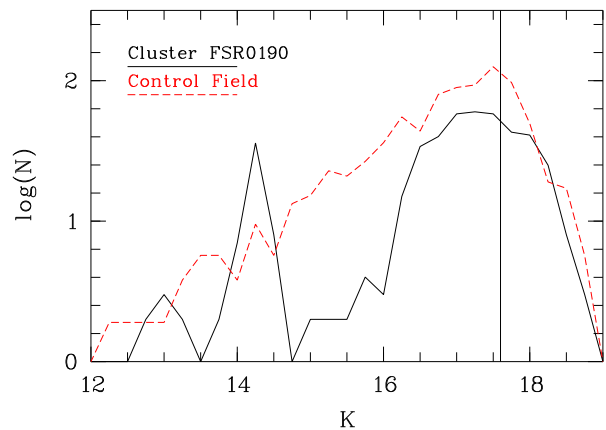}

\caption{\label{klum} K-band luminosity function (solid line) of the
decontaminated cluster stars. As dashed line we show the K-band luminosity
function of the control field, normalised to the same area. The vertical line
indicates the K-band completeness limit.} 

\end{figure}

The cluster distance converts to a distance of $R_{\rm GC}$\,=\,10.5\,kpc from
the Galactic centre (assuming R$_\odot$\,=\,7.2\,kpc, Bica et al.
\cite{2006A&A...450..105B}). The core radius of the cluster is then $r_{\rm
core}$\,=\,3.2\,pc. In Fig.\,\ref{klum} we show the K-band luminosity function
of the cluster area after decontamination, compared to the scaled control field.
As already discussed with the nearest neighbour plots, the field star density is
slightly higher than the average star density in the cluster. The number of
stars in the core helium burning stage is about 50 (slightly depending on the
decontamination). This can be used to estimate the total mass of the cluster.
According to Salaris \& Girardi \cite{2002MNRAS.337..332S} this corresponds to
10$\pm$3\,$\cdot$\,10$^4$M$_\odot$ for a population of stars with the age
and metallicity range of FSR\,0190. We also can integrate the brightness of all
detected cluster stars to $M_{\rm K}$\,=\,$-$6.6\,mag or $M_{\rm
V}$\,=\,$-$4.7\,mag (using V-K\,=\,1.9, Leitherer et al.
\cite{1999ApJS..123....3L}). Note that the mass and absolute brightness
estimates should be considered lower limits, since a number of potential cluster
RGB/AGB stars has been excluded from our analysis due to insufficient quality in
the photometry. Up to 40\,\% of the stars in the cluster RGB region are
affected, potentially almost doubling the mass estimate and increasing the
integrated brightness by 0.5\,mag. 

\section{Discussion and Conclusions}

\label{discussion}

\begin{figure}
\centering
\includegraphics[width=6cm, angle=-90]{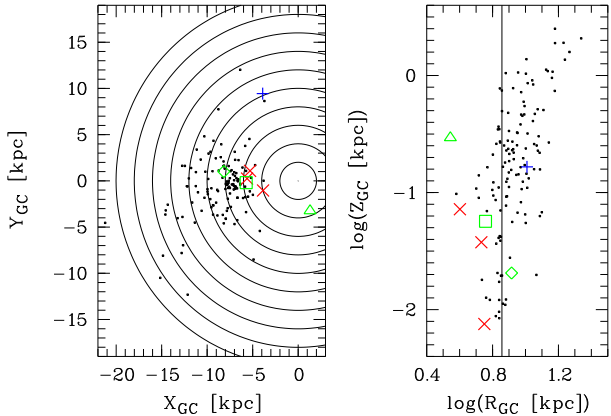}

\caption{\label{old} {\bf Left:} Distribution of old (age $\ge$ 1\,Gyr) open
clusters from the WEBDA database (dots) in the Galactic Plane. Also shown are
the so far identified old star clusters from the sample of Froebrich et al.
(2007b) -- crosses, FSR\,0031/0089/1744 (Bonatto \& Bica 2007b); triangle,
FSR\,1735 (GlCl, Froebrich et al. 2007a); square, FSR\,1767 (GlCl, Bonatto et
al. 2007); lozenge, FSR\,0584 (Bica et al. 2007); $+$ sign, FSR\,0190, this
work. Circles indicate distances from the Galactic Centre in steps of 2\,kpc.
{\bf Right:} Distance ($z$) to the Galactic plane of the same objects vs. the
distance to the Galactic Centre. The vertical line indicates the  Galactocentric
distance of the Sun.} 

\end{figure} 

\begin{table}
\centering
\caption{\label{properties} Measured properties of the cluster FSR\,0190}
\begin{tabular}{lr}
Parameter & Value  			\\
\hline
R.A. (J2000) & 20 05 31.3 	                    \\
DEC (J2000) & $+$33 34 09 	                    \\
$l$ [deg] & 70.7302 		                    \\
$b$ [deg] & $+$0.9498 		                    \\
$A_{\rm K}$ [mag] & 0.8$\pm$0.1                     \\
Age [Gyr] & $>$7		                    \\
$[$M/H$]$ & $-$0.9$\pm$0.4 	                    \\
$r_\odot$ [kpc] & 10.0$\pm$1.0                      \\
$R_{\rm GC}$ [kpc] & 10.5$\pm$0.8                   \\
$z$ [pc] & 170$\pm$15                               \\
$r_{\rm core}$ [pc] & 3.2$\pm$0.4                   \\
$r_{\rm tid}$ [pc] & 9.7$\pm$2.0                    \\
$M_{\rm K}$ [mag] & $-$6.6$\pm$0.2                  \\
M$_{\rm cl}$ [10$^4$\,M$_\odot$] & 10$\pm$1         \\
\end{tabular}
\end{table}

We have summarised the determined properties of the cluster FSR\,0190 in
Table\,\ref{properties}. Despite the significant contamination with field stars
we can classify FSR\,0190 as an old cluster with an age of more than 7\,Gyr. In
Fig.\,\ref{old} we compare the clusters position in the Galactic plane with the
other known old (age $>$\,1\,Gyr) Galactic clusters taken from the
WEBDA\footnote{http://www.univie.ac.at/webda/} database. Also plotted in this
diagram are the so far verified old clusters from the sample of Froebrich et al.
\cite{2007MNRAS.374..399F} (crosses, FSR\,0031/0089/1744, Bonatto \& Bica
\cite{2007MNRAS.in.pressB}); triangle, FSR\,1735, GlCl, Froebrich et al.
\cite{2007MNRAS.377L..54F}; square, FSR\,1767, GlCl, Bonatto et al.
\cite{2007MNRAS.in.pressB2}; lozenge, FSR\,0584, GlCl, Bica et al.
\cite{2007A&A...472..483B}). The figure shows, that FSR\,0190 might be one of
the most distant known old open clusters in the Milky Way. There are only a few
other known clusters with comparable ages and distances. The cluster also nicely
follows the relation of distance $z$ to the Galactic Plane with $R_{\rm GC}$, as
can be seen in the right panel of Fig.\,\ref{old}. Note that FSR\,1735 does not
follow the relation in this plot, in accordance to its GlCl nature. 

Could FSR\,0190 also be another so far unknown GlCl of the Milky Way? There are
obviously no definite arguments against such an interpretation. According to
Fig.\,\ref{iso} an age above 8\,Gyr and a metallicity below [M/H]\,=\,-0.75\,dex
(the most common values for the galactic GlCl sample -- Harris
\cite{1996AJ....112.1487H}) are certainly possible for this object. The
integrated luminosity of $M_{\rm V}$\,=\,$-$4.7\,mag, or slightly brighter,
would place it at the fainter end of the GlCl distribution, well within the
Palomar-like GlCl regime. This type of object might be considerably more
abundant given the number of recent discoveries (e.g. Bonatto et al.
\cite{2007MNRAS.in.pressB2} and Bica et al. \cite{2007A&A...472..483B}).
However, its position in the Galaxy nicely follows the relation for old Galactic
clusters, and its optical appearance does not show a centrally condensed
cluster. It is hence not possible to be absolutely certain that FSR\,0190 is a
GlCl. Nevertheless, such a possibility can currently not be excluded.

With the identification of FSR\,0190 as a distant old Milky Way cluster, now
seven objects in the sample of Froebrich et al. \cite{2007MNRAS.374..399F} have
been confirmed as so far unknown old Milky Way clusters. The distribution of
these objects shows that the FSR-sample contains a number of clusters in the
least complete regions, i.e. the inner Galaxy and/or distant objects. A
continued effort to classify the other cluster candidates will further enhance
the completeness of the known cluster sample. This will greatly improve studies
of star cluster mortality, as well as the formation history of the Milky Way.

\section*{acknowledgments}

We would like to thank the referee Sergio Ortolani for helpful comments to
improve the analysis of our data. The United Kingdom Infrared Telescope is
operated by the Joint Astronomy Centre on behalf of the Science and Technology
Facilities Council of the U.K.  The data reported here were obtained as part of
the UKIRT Service Program. This research has made use of the WEBDA database,
operated at the Institute for Astronomy of the University of Vienna.

\label{lastpage}

\end{document}